\documentclass[prl, twocolumn, aps, superscriptaddress,]{revtex4-2}
\usepackage{stackengine}
\usepackage{empheq}
\usepackage{amsmath} 
\usepackage{braket}
\usepackage{amsfonts}
\usepackage{amssymb}
\usepackage{graphicx} 
\usepackage{float}
\usepackage{bm}
\usepackage{color}
\usepackage[normalem]{ulem}
\usepackage[pdfencoding=auto]{hyperref}
\usepackage{lipsum}

\graphicspath{{.}{Figs/}}

\definecolor{darkgreen}{rgb}{0, 0.5, 0.05}

\newcommand{\deleted}[1]{}

\newcommand{\eqn}[1]{\begin{align}#1\end{align}}
\newcommand{\bs}[1]{\boldsymbol{#1}}
\newcommand{\ps}[1]{\partial_{#1}}
\newcommand{\pare}[1]{\left( #1 \right) }
\newcommand{\corchete}[1]{\left[ #1 \right]}
\newcommand{\fr}[2]{\frac{#1}{#2}}

\newcommand{\avg}[1]{\langle #1 \rangle}

\def\dd{\mathrm{d}}  

\def\bna{\bs{\nabla}}

\def\btau{\bs \tau}

\def\bu{\bs{u}}
\def\bv{\bs{v}}

\def\bbf{\bs{f}}

\def\bg{\bs{g}}

\def\bJ{\bs{J}}

\begin{document}

\title{Concentration Matters: Enhancing Particle Settling in Narrow Tilted Channels}
\date{\today}

\begin{abstract}

  Particles are known to sediment faster in containers with tilted walls than in vertical ones, a phenomenon known as the Boycott effect.
  In this work, we investigate how the tilt angle influences sedimentation in narrow channels across different particle volume fractions.
  Using particle-resolved computational fluid dynamics simulations, we reveal that there exists a concentration-dependent optimal tilt angle that maximizes sedimentation rates.
  Furthermore, at large tilt angles, the flow profiles across the channel deviate from the classical parabolic shape.
  We show that these non-parabolic profiles can be accurately captured by a one-dimensional Brinkman model,
  providing a predictive framework for understanding and tuning sedimentation in tilted geometries.
  Our findings demonstrate the potential to control and optimize particle settling by adjusting the channel tilt according to particle concentration,
  opening new possibilities for design in industrial and laboratory processes.
  
\end{abstract}
 
\author{Dipankar Kundu}
\affiliation{BCAM - Basque Center for Applied Mathematics, Mazarredo 14, Bilbao, E48009, Basque Country - Spain}

\author{Florencio Balboa Usabiaga}
\affiliation{BCAM - Basque Center for Applied Mathematics, Mazarredo 14, Bilbao, E48009, Basque Country - Spain}

\author{Adolfo Vázquez-Quesada}
\affiliation{Departamento de Física Fundamental, UNED, Apartado 60141, Madrid, 28080, Spain}

\author{Marco Ellero}
\affiliation{BCAM - Basque Center for Applied Mathematics, Mazarredo 14, Bilbao, E48009, Basque Country - Spain}
\affiliation{IKERBASQUE, Basque Foundation for Science, Calle de Maria Díaz de Haro 3, 48013, Bilbao, Spain}
\affiliation{Complex Fluids Research Group, Department of Chemical Engineering, Faculty of Science and Engineering, Swansea University, Swansea SA1 8EN, United Kingdom}

\maketitle

The separation of particles by sedimentation is a seemingly simple process that hides many interesting phenomena
\cite{Segre1997, Heitkam2013, Goldfriend2017, Sprinkle2021}.
For an isolated particle, the sedimentation velocity results from the balance between gravitational and drag forces \cite{Happel1983}.
As the particle concentration increases, the average sedimentation velocity decreases due to the counterflow induced by the motion of neighboring particles
severely, increasing sedimentation times \cite{Batchelor1972, Brady1988b, Kundu2025}.
This reduction depends strongly on the particle volume fraction, $\phi$.
For instance, the widely used empirical Richardson–Zaki law predicts a sedimentation velocity of
$u(\phi) = u_0 (1 - \phi)^n$,
where $u_0$ is the single-particle sedimenting velocity and $n \sim 5$ \cite{Richardson1997}.
Consequently, strategies to enhance sedimentation in dense suspensions are of great practical and engineering relevance \cite{Reyes2022}.

\begin{figure*}
  \centering
  \includegraphics[width=0.7\textwidth]{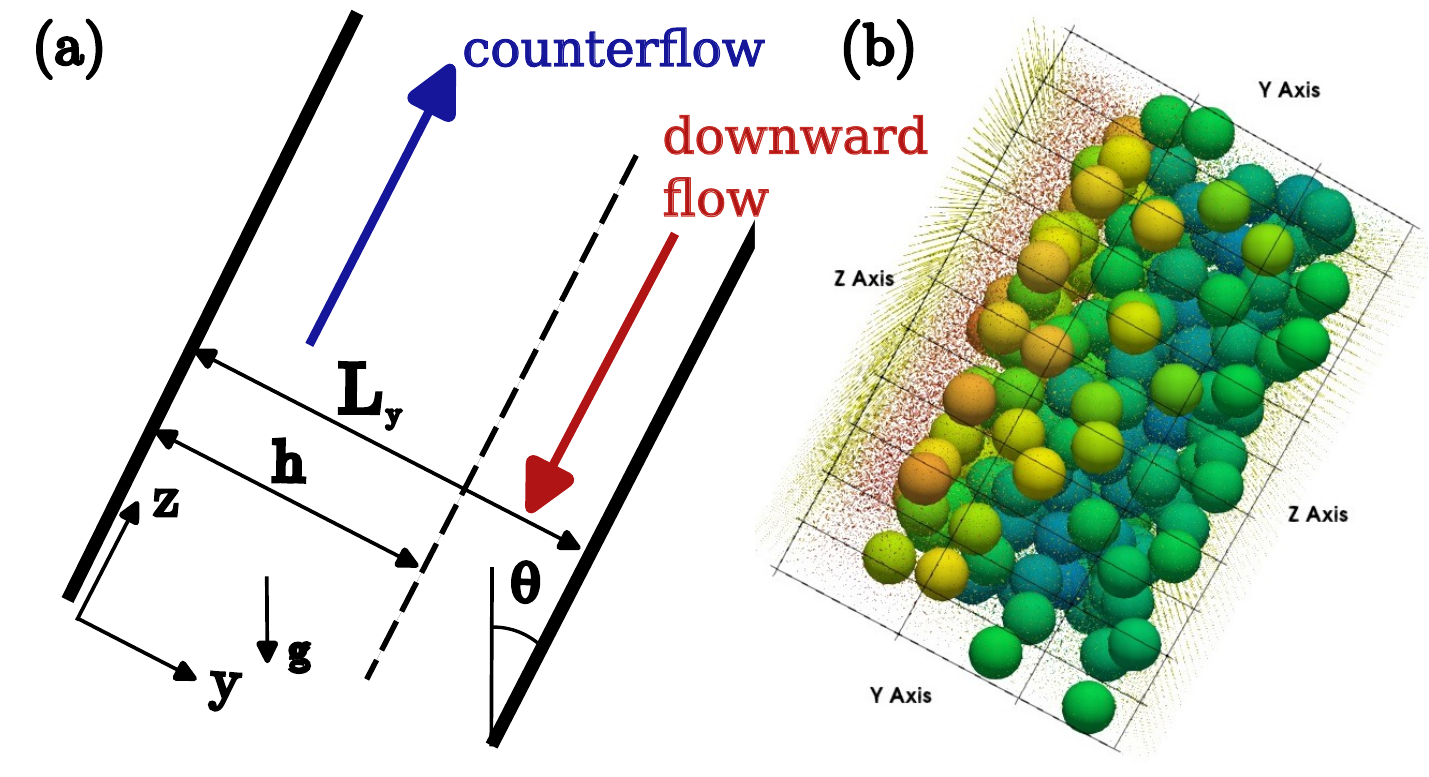}
  \caption{{\bf (a)} Schematic view of the domain geometry, a channel of span $L_y$ tilted an angle $\theta$ respect gravity $\bg$.
    The solute particles sediment primarily near the lower wall while a counterflow is generated near the upper wall.   
    The dashed line represents the interface between the clear fluid on the top and the particle suspension on the bottom of the channel.
    {\bf (b)} Snapshot of a SPH simulation at volume fraction $\phi=0.2$ and tilt angle $\theta=30^{\circ}$.
    The particles are represented as solid spheres while the dots represent SPH particles.}
  \label{fig:sketch}
\end{figure*}

A very popular enhancement mechanism is related to the so-called \emph{Boycott effect}, in which suspensions sediment faster in long,
tilted channels (or tubes) than in simple vertical ones \cite{Boycott1920}.
This counterintuitive phenomenon can be cleared up by two complementary explanations, one geometrical and one hydrodynamic.
The geometrical argument stems from the larger projected area of tilted channels across which particles can settle.
This geometric contribution has been captured by the kinetic Ponder-Nakamura-Kuroda (PNK) model,
often employed to fit experimental data \cite{Ponder1925, Peacock2005, Guazzelli2011}.
The hydrodynamic explanation, on the other hand, arises because particles tend to sediment near the lower wall,
while the compensating upward flow develops primarily near the upper wall (see Fig.\ \ref{fig:sketch}) \cite{Acrivos1979, Buerger2012, Heitkam2013, Chang2019, Zeng2021}.
Because the particles settling close to the lower wall are largely unaffected by the counterflow, they sediment faster than in vertical channels.
Thus, by breaking the system symmetry in the horizontal direction, 
the Boycott effect provides an effective means to circumvent the sedimentation slowdown linked to the Richardson–Zaki law, 
and, as consequence of its simplicity,
it has found applications across a wide range of engineering systems \cite{Culp1968, Kim2013, HincapieGomez2015, Reyes2022}.

In this Letter we investigate a surprisingly overlooked aspect of the Boycott effect,
and demonstrate the existence of an optimal tilt angle to enhance sedimentation.
Moreover, we show that this angle varies with the initial particle volume fraction.
We show quantitatively  that the suspension can be modeled as a Brinkman fluid,
which enables us to develop a theory predicting $\phi$-dependent optimal angles.
This finding can serve to optimize separation problems targeting specific systems.

The problem can be formulated as follows, we consider a system of $N$ spherical particles sedimenting in a tilted channel as depicted in Fig.\ \ref{fig:sketch}.
The walls are perpendicular to the $y$-direction, and the system is periodic along the $x$-$z$ directions.
The periodic boundary conditions allow us to study steady-state sedimentation as if the channel were much longer,
without increasing the computational cost.

The solvent obeys the Navier--Stokes equations
\eqn{
  \label{eq:NS}
  \bna \cdot \bv &= 0, \\
  \rho \pare{\ps{t} \bv + \bv \cdot \bna \bv} &= -\bna p + \eta \bna^2 \bv + \bbf^{Q}(t),
}
which represent the conservation of mass and momentum.
In the above equations, $\bv$ and $p$ are the flow velocity and pressure respectively, and $\eta$ is the solvent shear viscosity.
The last term in the momentum equation, $\bbf^{Q}(t)$, is a density force that ensures zero mass flux along the $z$-axis, 
$\label{eq:Q}  Q = \int v_z \, \dd x \dd y = 0,$
where the integral runs over both the fluid and particle domains.

The dynamics of the suspended particles of mass $m$ are governed by the Newton--Euler equations
\eqn{
  \label{eq:newton}
  m \fr{\dd \bu_{\alpha}}{\dd t} = \bbf_{\alpha}, \;\;\;
  \fr{\dd \bJ_{\alpha}}{\dd t} = \btau_{\alpha},
}
where $\bu_{\alpha}$ and $\bJ_{\alpha}$ are the velocity and angular momentum of particle $\alpha$,
and $\bbf_{\alpha}$ and $\btau_{\alpha}$ the total force and torque acting on the particle. 
Here, we assume the torque is purely hydrodynamic, while the force includes hydrodynamic, steric, and gravitational contributions. 
The hydrodynamic forces are computed by enforcing the no-slip boundary condition of the liquid at the particle surfaces. 
The repulsive interaction between the solid particles and between the solid particles and the walls prevents overlaps while 
the gravitational interaction is given by a constant force on the particles, aligned with the vertical direction 
\eqn{
  \label{eq:g_force}
  \bbf_{\alpha}^{\text{g}} = \fr{4\pi}{3} a^3 \rho g \pare{0, \sin \theta, -\cos \theta}.
}
As shown in Fig.\ \ref{fig:sketch}, the $z$-axis is aligned with the channel, and thus the direction of the gravitational force depends on the tilt angle $\theta$.

To solve the Navier--Stokes equations and the fluid--structure
interactions, we employ the Smoothed Particle Hydrodynamics (SPH)
method that has been used for modeling particle suspensions (see details in Supplementary Sec.\ I)
\cite{Bian2012, Vazquez-Quesada2016, Vazquez-Quesada2019}.
In this work, we use the particle radius $a=1$ as the unit length,
the sedimentation time of an isolated particle as the unit time $\tau = 9 \eta / (2 a \rho g) = 1$
and the density is set to $\rho=1$.
We fix the channel dimensions to $L_x \times L_y \times L_z = 22 \times 14 \times 22$ and
we vary the tilt angle $\theta$ between $0^{\circ}$ (vertical channel) and $90^{\circ}$ (horizontal channel),
as well as the number of particles $N$, to sample solid volume fractions $\phi$ in the range $\phi \sim 0.1 \mbox{--} 0.4$.
The viscosity and gravitational force are set such that the Reynolds number is negligible, $\text{Re} = 0.016 \ll 1$
(see the full parameter list in Supplementary Table~I).

\begin{figure}
  \centering
  \includegraphics[width=0.9\columnwidth]{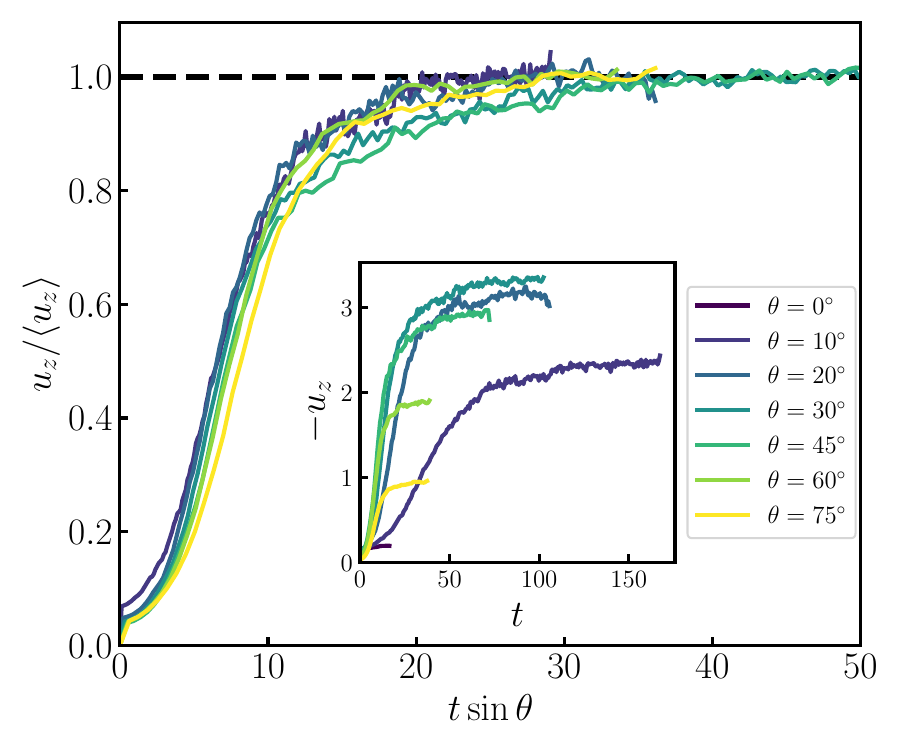}
  \caption{Particle sedimentation velocities versus time for volume fraction $\phi=0.3$ and tilts
    $\theta \in [0^{\circ}, 75^{\circ}]$ (curves from dark to light).
    In the main figure the velocities and the time are scaled with the tilt-dependent steady state velocity
    $\avg{u_z}$ and with $1/\sin \theta$ respectively.
    Inset: same data without scaling.
    The curve for $\theta = 0^{\circ}$ is only shown in the inset.
    Results for other volume fractions are shown in the Supplemental Fig.\ 1.
  }
  \label{fig:vel_vs_time}
\end{figure}

We start all simulations with the particles arranged randomly across the channel and the system at rest and
run the simulations for dozens of sedimentation times until the system reaches a steady state.
To guarantee that the system is in the steady state,
we measure the particle-averaged sedimentation velocity, $u_z(t) = \fr{1}{N}  \sum_{\alpha=1}^N u_{z,\alpha}(t)$,
from the last fourth of the simulation and verify that its relative standard deviation over time is below 3\% in all cases.
In vertical channels ($\theta=0^{\circ}$), the particles attain their steady state sedimentation velocity
in a few characteristic viscous times, $\tau_{\text{vis}} = \rho a^2 / \eta \sim 0.02$.
In tilted channels, however, the system takes much longer to reach the steady state, see inset in Fig.\ \ref{fig:vel_vs_time}.
The longer relaxation times are explained by the rearrangement of the particles in the $y$-direction;
this rearrangement time is inversely proportional to the gravitational force along the $y$-axis.
Thus, for a given volume fraction and channel width, the velocity-vs-time curves collapse when
time is scaled with $1 / \sin \theta$ and velocity with the steady state velocity $\avg{u_z}$, as shown in the main panel of Fig.\ \ref{fig:vel_vs_time}.

\begin{figure}
  \centering
  \includegraphics[width=0.9\columnwidth]{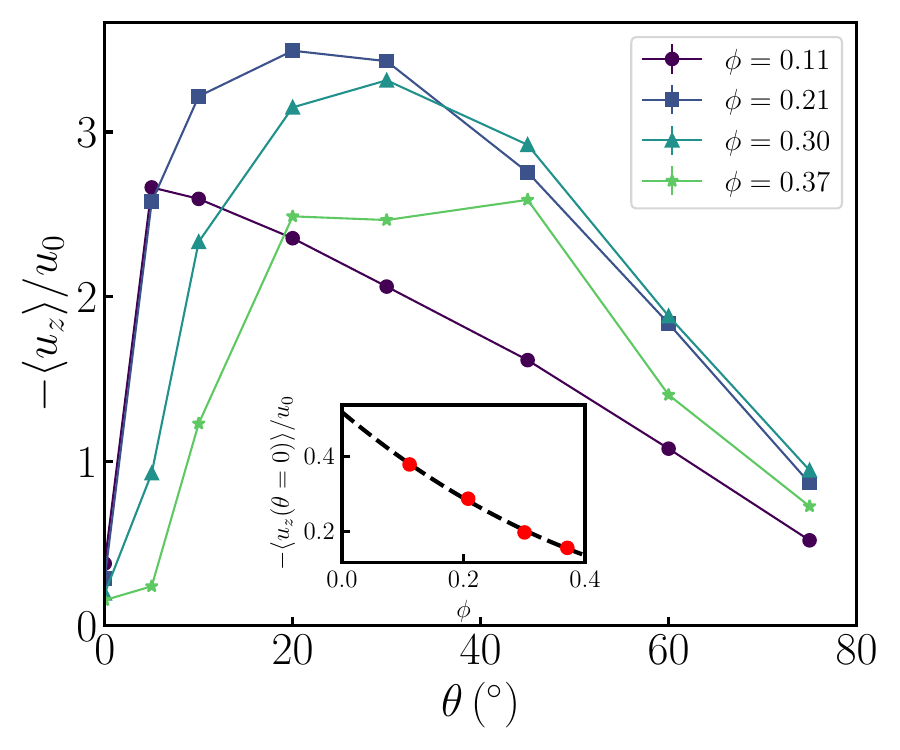}
  \caption{Steady state sedimentation velocity normalized with the single-particle sedimenting velocity, $u_0$,
    versus tilt angle for different volume fractions.
    The inset shows the steady state sedimentation velocity in vertical channels versus the volume fraction and
    the fit $-\avg{u_z} \sim (1-\phi)^n$ with $n \approx 2.6$.
  }
  \label{fig:vel_vs_tilt}
\end{figure}

The change of steady-state velocity with the tilt angle is shown, for several volume fractions, in Fig.\ \ref{fig:vel_vs_tilt}.
First, we can observe that in vertical channels ($\theta=0^{\circ}$), the sedimentation velocity decreases with the volume fraction.
The results are well fitted by the Richardson-Zaki law with a smaller exponent ($n \approx 2.6$) than in bulk as previously reported
for particles in narrow channels \cite{Schwarzer1995, Kuusela2004}.
More interesting, a positive tilt increases the sedimentation velocity beyond the average velocity of an isolated particle for all volume fractions considered.
The speed enhancement with respect to the vertical channel is quite dramatic, from a factor of seven for $\phi=0.11$ up to a factor of sixteen for $\phi=0.3$.
This sedimentation speed-up is precisely related to the hydrodynamic contribution to the Boycott effect \cite{Acrivos1979, Buerger2012, Heitkam2013, Chang2019, Zeng2021}.
Strikingly, the optimal angle to maximize sedimentation is concentration dependent, as deduced from the steady-state sedimentation
velocity curves shown in Fig.\ \ref{fig:vel_vs_tilt}.


\begin{figure*}
  \centering
  \includegraphics[width=0.3\textwidth]{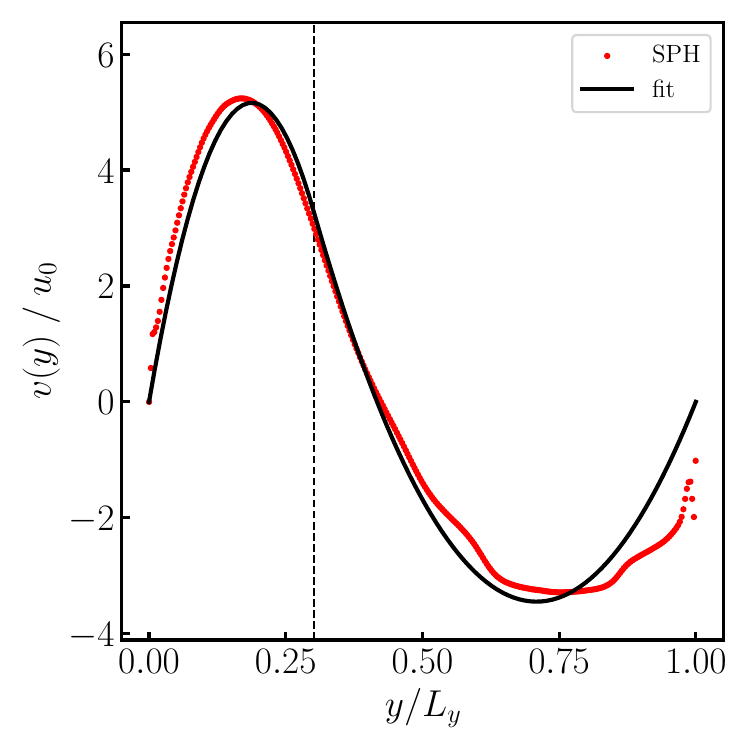}
  \includegraphics[width=0.3\textwidth]{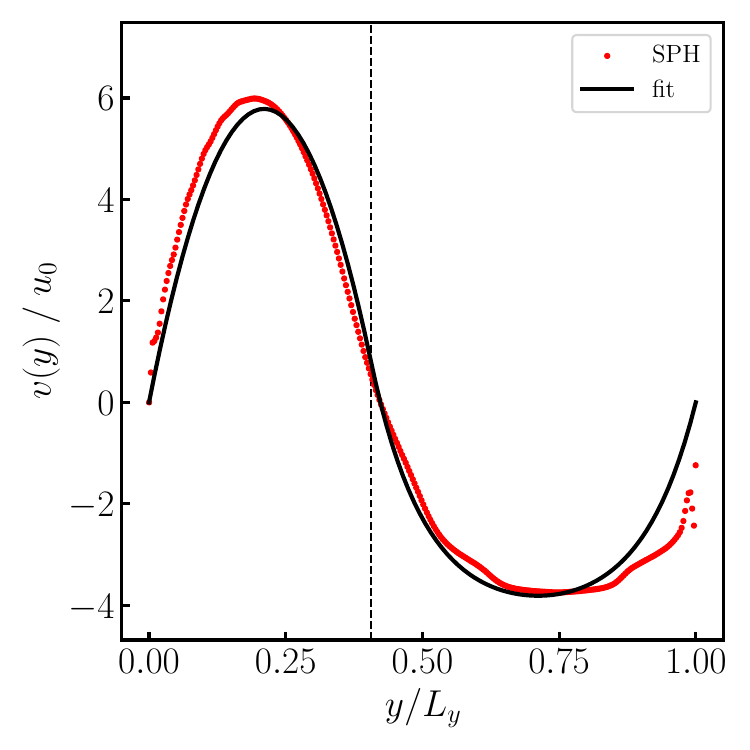}    
  \includegraphics[width=0.3\textwidth]{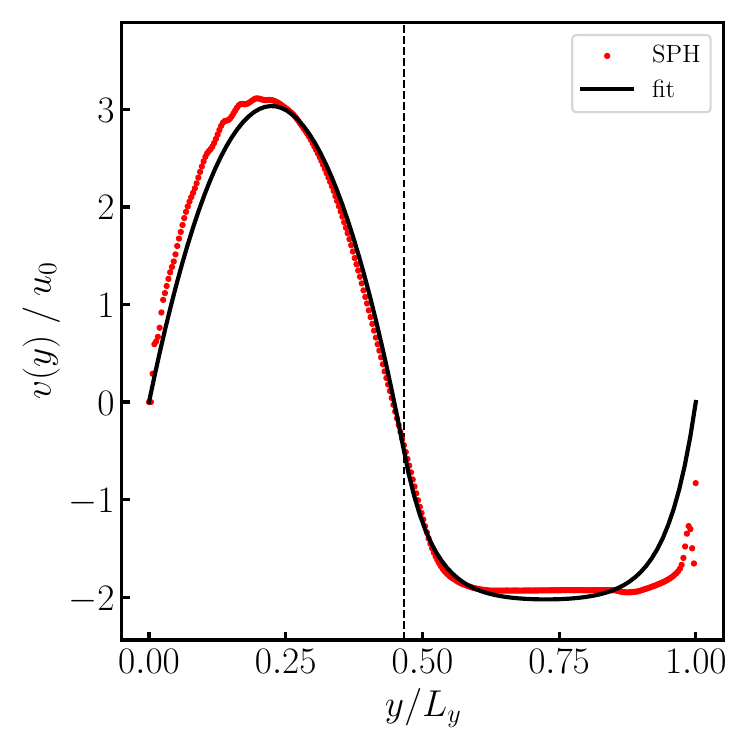}  
  \caption{Velocity profiles across the channel for volume fraction $\phi=0.3$ and tilt angles (from left to right) $\theta=10,\, 20 \text{ and } 60^{\circ}$.
    The velocity is normalized with the isolated particle velocity $u_0=1$.
    The dashed vertical line represents the interface position, $h$, obtained from the fits to the one-dimensional model,
    Eqs.\ \eqref{eq:model_stokes}-\eqref{eq:model_up}.
    Results for other volume fractions are shown in the Supplemental Fig.\ 2.
  }
  \label{fig:profile}
\end{figure*}

To better characterize the steady-state sedimentation, we compute the suspension velocity profiles across the channel (see Fig.\ \ref{fig:profile}). 
The profiles are divided into two main domains: one near the upper wall, associated with the clear fluid, where the fluid flows upward, and another 
near the lower wall, associated with the particulate suspension flowing downward. 
The clear-fluid part of the profile is always parabolic, as expected for a Newtonian flow in a channel. 
Interestingly, the particulate-suspension part of the profile adopts a parabolic shape for low tilt angles, 
but it flattens strongly for larger values of $\theta$ 
reflecting the non-Newtonian character of dense suspensions (see Fig.\ \ref{fig:profile} and Supplementary Fig.\ 2).

In order to qualitative understand the system's behavior,
we derive a one-dimensional continuous model that reproduces the velocity profiles.
In this model, we assume the flow to be divided into two domains separated by an interface at $h$ as depicted in the sketch in Fig.\ \ref{fig:sketch}a.
The first, a clear fluid near the upper wall that we model as a Newtonian fluid with viscosity $\eta$;
the second, a suspension near the lower wall which is modeled as a Brinkman fluid with the same viscosity $\eta$ and a permeability $\kappa^2$,
where $\kappa$ has units of length \cite{Brinkman1949, Housiadas2014}, and subject to a density force along the channel.
In this model, we assume that all the motion occurs along the channel and that the velocity only changes along $y$.
Thus, the Stokes and Brinkman equations are
\begin{alignat}{2}
  \label{eq:model_stokes}
  -\ps{z} p + \eta \ps{y}^2 v = 0  \;\; y \in [0, h], \\
  \label{eq:model_brinkman}  
  -\ps{z} p + \eta \ps{y}^2 v  - \fr{\eta}{\kappa^2} (v - u_p)  = f_g \cos \theta  \;\; y \in [h, L_y], \\
  \label{eq:model_div}  
  \ps{z} v = 0  \;\; y \in [0, L_y], \\
  \label{eq:model_up}    
  u_p = \fr{1}{L_y - h} \int_h^{L_y} v \, \dd y - \fr{2 \rho g a^2 \cos \theta}{9 \eta}. 
\end{alignat}
The parameter $u_p$, appearing in the Brinkman equation \eqref{eq:model_brinkman}, represents the velocity of the solid fraction as it sediments along the channel.
The model assumes that the solid fraction is advected by the flow while it sediments with the velocity of an
isolated particle, first and second terms in the right hand side of Eq.\ \eqref{eq:model_up}.
The magnitude of the density force in Eq.\ \eqref{eq:model_brinkman}, $f_g = \phi \rho g L_y / (L_y - h) - 2\rho g a^2/ (9 \kappa^2)$,
ensures that the total force in the system is the same as the gravity force in the SPH simulations.

The model is closed with the no-slip boundary conditions on the walls and the continuity of the velocity and its derivative on the interface
\eqn{
  v(0) = v(L_y) = 0, \\
  v(h^-) = v(h^+), \\
  \ps{y} v(y)\vert_{y=h^-} = \ps{y} v(y)\vert_{y=h^+}.
}
Additionally, as in the SPH simulations, we enforce a zero total flux of mass 
\eqn{
  \label{eq:Q_v2}
  Q = \int_0^{L_y} v(y)\, \dd y = 0.
}
In the clear fluid domain the general solution is parabolic $v(y) = v_1 y + v_2 y^2$ 
while in the Brinkman domain the profile is $v(y) = v_3 \corchete{\exp\pare{(y-L_y)/\kappa} - 1} +  v_4 \corchete{\exp\pare{-(y-L_y)/\kappa} - 1}$. 
The coefficients of the general solution are functions of the model parameters, e.g.\ $v_1 = v_1(L_y,\eta,g,a, \phi, \theta, h, \kappa)$. 
We fit the general solution to the numerical profiles using $\kappa$ and $h$ as a fitting parameters, 
while the others are set to the values used in the SPH simulations.

The fits shown in Fig.\ \ref{fig:profile} demonstrate that the one-dimensional model can capture both the parabolic and flat profiles obtained numerically.
The velocity of the solid fraction, $u_p$, is well correlated with the steady sedimentation velocity measured directly from the SPH simulations,
i.e.\ $u_p \approx \avg{u_z}$, see Supplementary Fig.\ 3.
This agreement supports the use of a Brinkman fluid to model the sedimenting particulate suspension.
The permeability, $\kappa^2$, decreases with the effective volume fraction of the suspension near the lower wall,
defined as $\phi_{\text{eff}} = \phi L_y / (L_y - h)$, see Supplementary Fig.\ 3.
However, the values of $\kappa$ do not collapse to a single curve, thus, the suspension is not completely characterized by this one-dimensional model.

\begin{figure}
  \centering
  \includegraphics[width=0.9\columnwidth]{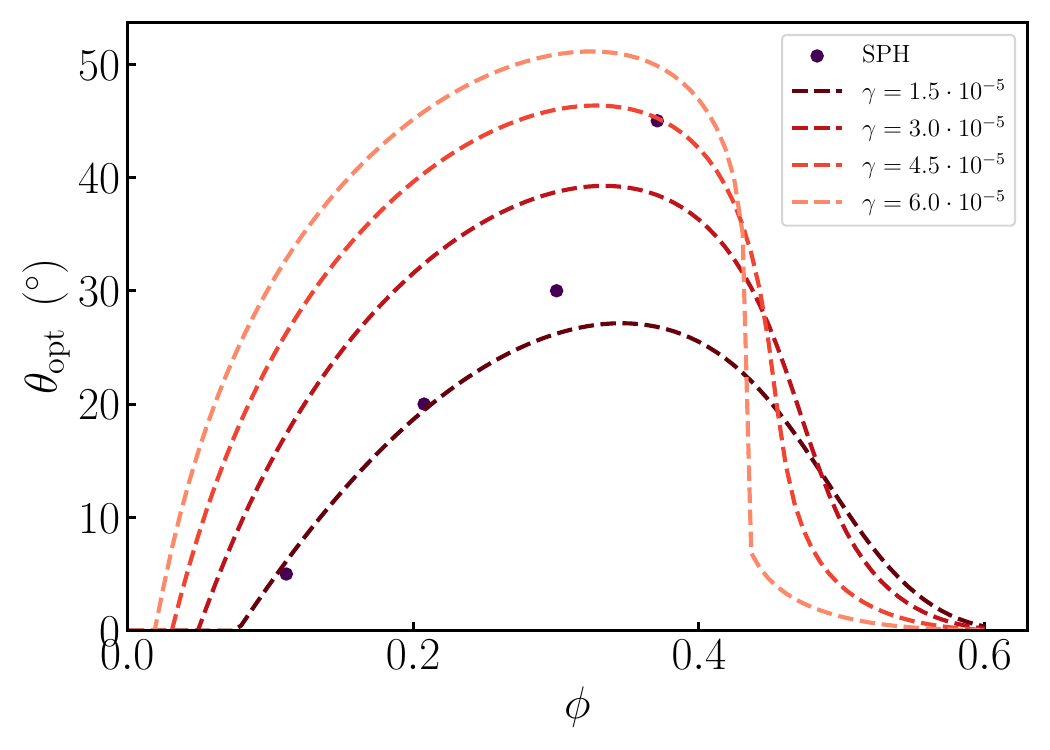}
  \caption{Optimal angle to enhance sedimentation, $\theta_{\text{opt}}$, versus volume concentration
    obtained from the SPH simulations (dots) and the one-dimensional model (lines).
    Theoretical predictions for $\gamma = 1.5 \cdot 10^{-5},\, 3 \cdot 10^{-5},\, 4.5 \cdot 10^{-5} \text{ and } 6 \cdot 10^{-5}$
    from dark to light colors.
  }
  \label{fig:opt}
\end{figure}

Next, we outline how the one-dimensional model can be used to predict the optimal sedimentation angle, $\theta_{\text{opt}}$,
that maximizes the sedimentation velocity for a given $\phi$.
The model predicts a sedimentation velocity of the form
$u_p = U(L_y, \eta, g, a, \phi, h, \kappa) \cos\theta$,
where, in general, $h=h(\phi,\theta)$ and $\kappa=\kappa(\phi,\theta)$.
In the model, the position of the interface depends on the effective volume fraction near the wall
as $h = (1 - \phi / \phi_{\text{eff}}) L_y$.
The effective volume fraction increases with the tilt angle as gravity compacts the particles in the $y$-direction.
Inspired by the suspension model of Wyart and Cates \cite{Wyart2014},
we model the effective concentration as $\phi_{\text{eff}} = \phi + (\phi_{\text{max}} - \phi)f(\theta)$,
where $\phi_{\text{max}} = 0.64$ is near the jamming fraction and $f(\theta)$ is a monotonically decreasing function.
For simplicity we use $f(\theta) = \exp[-\gamma(\theta - 90^{\circ})^2]$, where the parameter $\gamma$ controls how steep the transition towards $\phi_{\text{max}}$ is.
Numerical evaluation reveals that the results are less sensitive to the values of $\kappa$ than to those of $h$ over the range of values obtained in the fits,
thus, we set $\kappa = 1$.
In Fig.\ \ref{fig:opt} we plot the optimal angle versus $\phi$ predicted by the model for several values of $\gamma$,
together with the optimal angle obtained from the SPH simulations.
The exact shape of the curves vary with the models for $h$ and $\kappa$,
but, quite generally, the optimal angle tends to zero for $\phi \rightarrow 0$,
takes finite values for intermediate volume fractions, and finally, tends to zero when the volume fraction approaches $\phi_{\text{max}}$.
This last result is expected; if the flow cannot be segregated into downward and upward streams because the particles
occupy the whole channel, the sedimentation speed $u_p$ is maximized when the gravitational force is aligned with the channel.

The present model enables a more precise characterization and control of sedimentation.
For instance, if one can tune the volume fraction and wishes to maximize the mass flux, $j_m = \rho \phi u_p$ (see Supplementary Fig.\ 4),
the one-dimensional model can be used to identify the values of $\phi$ and $\theta$ that maximize $j_m$.

In conclusion, we have demonstrated the existence of an optimal tilt angle that enhances sedimentation and that this angle depends on the particle volume fraction.
We have shown that the time required to reach the steady state scales inversely with the gravitational forcing along the transverse direction, i.e.\ as $\sim 1 / \sin\theta$.
Moreover, we find that $\theta_{\text{opt}} \rightarrow 0$ both in the dilute limit $\phi \rightarrow 0$ and near close packing $\phi \rightarrow \phi_{\text{max}}$,
whereas a nonzero optimal angle emerges for intermediate values of $\phi$.
This insight provides a practical guideline for improving the efficiency of sedimentation devices without requiring full knowledge of the underlying suspension properties.
Finally, we show that the system can be faithfully described by a one-dimensional Stokes–Brinkman model with only two free parameters, $\kappa$ and $h$,
offering a simple yet powerful framework to predict and control sedimentation behavior.

\section*{Acknowledgments}
We thank Dariel Hernández for useful discussions.
This work has been partially funded by the Basque Government through the projects Elkartek SosIAMet KK-2022/00110 and the BERC 2022-- 2025 program.
Financial support by the Spanish State Research Agency through BCAM Severo Ochoa Excellence Accreditation
CEX2021-001142-S/ MICIN/AEI/10.13039/501100011033 and the project
PID2024-158994OB-C42 (``Multiscale Modeling of Friction, Lubrication, and Viscoelasticity in Particle Suspensions'' and acronym ``MMFLVPS'')
funded by MICIU/AEI/10.13039/501100011033 and cofunded by the European Union. No.\ PID2020-117080RB-C55 and the project
No.\ PID2020-117807RB-C54 ``Coarse-Graining theory and experimental techniques for multiscale biological systems''.

\bibliographystyle{apsrev4-2}
\bibliography{Biblio}

\end{document}



\title{Concentration Matters: Enhancing Particle Settling in Narrow Tilted Channels}


\author{Dipankar Kundu}
\affiliation{BCAM - Basque Center for Applied Mathematics, Mazarredo 14, Bilbao, E48009, Basque Country - Spain}

\author{Florencio Balboa Usabiaga}
\email{fbalboa@bcamath.org}
\affiliation{BCAM - Basque Center for Applied Mathematics, Mazarredo 14, Bilbao, E48009, Basque Country - Spain}

\author{Adolfo Vázquez-Quesada}
\affiliation{Departamento de Física Fundamental, UNED, Apartado 60141, Madrid, 28080, Spain}

\author{Marco Ellero}
\affiliation{BCAM - Basque Center for Applied Mathematics, Mazarredo 14, Bilbao, E48009, Basque Country - Spain}
\affiliation{IKERBASQUE, Basque Foundation for Science, Calle de Maria Díaz de Haro 3, 48013, Bilbao, Spain}
\affiliation{Complex Fluids Research Group, Department of Chemical Engineering, Faculty of Science and Engineering, Swansea University, Swansea SA1 8EN, United Kingdom}


                    
\maketitle

\tableofcontents

\section{Numerical method}
\label{sec:method}
To  solve   the  Navier-Stokes   equations  and   the  fluid-structure
interactions,  we employ  the  Smoothed  Particle Hydrodynamics  (SPH)
method \cite{Monaghan1992, Hu2006, Bian2012, Vazquez-Quesada2016, Vazquez-Quesada2019}.
SPH    discretizes   the
Navier-Stokes  equations  using  a   Lagrangian  approach,  where  SPH
particles carry  mass and  momentum.  The discretized  conservation of
mass and momentum  read
\eqn{
  \label{eq:NS_sph_r}
  \br_i &= \bv_i, \\
  \label{eq:NS_sph_v}
  m_{\text{sph}} \dot{\bv}_i &= -\sum_j \corchete{\fr{p_i}{d_i^2} + \fr{p_j}{d_j^2}} \fr{\partial W(r_{ij})}{\partial r_{ij}} \be_{ij}
  + \sum_j \pare{D + 2} \eta \corchete{\fr{1}{d_i^2} + \fr{1}{d_j^2}} \fr{\partial W(r_{ij})}{\partial r_{ij}} \fr{\be_{ij} \cdot \bv_{ij}}{r_{ij}} \be_{ij} + \bbf^Q_i
}
where $\br_i$ and $\bv_i$ are the position and velocity of the SPH fluid particles, $m_{\text{sph}}$ is their mass and $D=3$ the space dimensionality.
In these equations, $W(r)$ is an interpolation kernel used to compute density and velocity gradients in the disordered mesh formed by the SPH particles.
Here we use the quintic kernel  with cutoff $r_{c,\text{sph}}=1.2$ \cite{morris1997}.

\begin{longtable}{llcc}
  \caption{Dimensionless parameters used in the present study: }\\  
  \hline
  Symbol & Description & Value & Comments \\ 
  \hline
  \endhead
  $a$ & Particle radius &  $1$ &  Unit length \\
  $\rho$ & Equilibrium fluid density & $1$ & Unit density \\
  $\tau_s$ & Sedimentation time & $1$ & Unit time, computed from the sedimentation time \\
  & & & of an isolated particle $\tau = \fr{9 \eta}{2 a \rho g}$ \\
  $L_y$ & Channel width & 14 & \\
  $L_x=L_z$ & Channel length & 22 & The system has PBC along the x and z axes \\
  $r_{c,\tex{sph}}$ & SPH cutoff radius &  $1.2$ \\
  $\phi$ & Solid volume fraction & $0.11,\, 0.21,\, 0.30 \text{ and } 0.37$   \\  
  $N$ & No.\ of solid particles & $180,\, 336,\, 486\, \text{ and } 600$ &  \\
  $N_{\text{sph}}$ & No.\ of SPH particles &  $\sim 10^5$ & the exact number varies slightly \\
  & & & with the volume fraction \\  
  $m_{\text{sph}}$ & SPH particle mass & 0.064 & \\
  $m$ & Particle mass & 4.189 & We employ the Boussinesq approximation \\
  & & & $m=4\pi \rho a^3 / 3$ \cite{Verzicco2023}. \\
  $\eta$ & Fluid shear viscosity & $64.41$  \\
  $\eta_b$ & Fluid bulk viscosity & $107.36$  \\
  $c$ & Speed of sound & $380.7$ \\
  $g$ & Gravity & $289.86$  \\
  $k_{\text{PID}}$ & Harmonic constant PI controller & 0.5 & \\
  $\tau_{\text{PID}}$ & Characteristic time PI controller & $6.57 \cdot 10^{-5}$ & $\tau_{\text{PID}} = 10 \dt$ \\     
  Re & Reynolds number & $0.016$ & Computed from the terminal velocity of an \\
  & & & isolated particle $\text{Re} = \fr{2 g}{9} \pare{\fr{a \rho}{\eta}}^2$ \\     
  Ma & Mach number & 0.0026 & Computed from the terminal velocity of an \\
  & & & isolated particle $\text{Ma} = \fr{2 \rho g a^2}{9 \eta c}$ \\
  $\dt$ & Time step size & $6.57 \cdot 10^{-6}$ & \\  
  \hline
  \label{tab:parameters}
\end{longtable}

The particle number density (or inverse SPH particle volume) is given by the sum of the interpolation kernel over its neighbors
\eqn{
  d_i = \fr{1}{\mcV_i} = \sum_j W(r_{ij}),
}
which defines the density of the SPH particles, $\rho_i = m_{\text{sph}} d_i$.
The SPH fluid is weakly compressible, to maintain the density approximately constant we employ the equation of state
\eqn{
  p_i = c^2 \left(\rho_i - \rho \right),
}
and we set the speed of sound large enough to make the Mach number negligible (see Table \ref{tab:parameters}).

The solid particles and the channel walls are discretized with SPH particles just like the fluid.
However, these particles are constrained to move as a rigid body in the case of the suspended particles,
or to remain fixed in the case of the walls \cite{Bian2012}.
The forces acting on the solid particles are decomposed into four terms
\eqn{
  \bbf_{\alpha} = \bbf^{\text{sph}}_{\alpha}  + \bbf^{\text{lub}}_{\alpha} + \bbf^{\text{g}}_{\alpha} + \bbf^{\text{rep}}_{\alpha}.
}
The first two are hydrodynamic forces, $\bbf^{\text{sph}}_{\alpha}$ is computed between the SPH particles forming the solid particles
and the fluid to impose the no-slip and no penetration conditions on the particle surfaces \cite{Bian2012}.
When two solid particles get too close for a given SPH resolution, the term $\bbf^{\text{sph}}_{\alpha}$ does not capture the hydrodynamic forces accurately.
To improve accuracy without dramatically increasing the computational cost, we include pairwise lubrication corrections, $\bbf^{\text{lub}}_{\alpha}$,
and we solve it implicitly as in Ref.\ \cite{Vazquez-Quesada2016a}.
The gravitational interaction is given by a constant force on the particles, aligned with the vertical direction
\eqn{
  \bbf_{\alpha}^{\text{g}} = \fr{4\pi}{3} a^3 \rho g \pare{0, \sin \theta, -\cos \theta}.
}
As shown in Fig.\ 1 in the main document,
the $z$-axis is aligned with the channel, and thus the direction of the gravitational force depends on the tilt angle $\theta$.
Finally, we include a repulsive interaction between the solid particles and between the solid particles and the walls,
$\bbf_{\alpha}^{\text{rep}} = \sum_\beta \bbf_{\alpha\beta}^{\text{rep}}$, where $\beta$ runs over all other particles and walls,
to avoid possible overlaps. The specific equation of the repulsive interaction is
\eqn{
  \boldsymbol{f}_{\alpha\beta}^{\text{rep}} = F_0 \frac{\tau e^{-s / \lambda}}{1-e^{-s / \lambda}}\boldsymbol{e}_{\alpha\beta},
}
where ${\be}_{\alpha\beta}$ is the unit vector between particles $\alpha$ and $\beta$,
$s$ is the dimensionless distance between the sphere-sphere or sphere-wall surfaces, $\lambda$ determines the interaction range and
$F^{\text{rep}}$ its magnitude (see Table \ref{tab:parameters}) \cite{sierou2002}
The torque on the particles, $\btau_{\alpha} = \btau_{\alpha}^{\text{sph}} +  \btau_{\alpha}^{\text{lub}}$, includes only hydrodynamic interactions
to impose the no-slip condition and to include the lubrication interactions between nearby particles.

The particle equations of motion, Eqs.\ 1--4 in the main document, and Eqs.\ \eqref{eq:NS_sph_r}-\eqref{eq:NS_sph_v}
are integrated using a semi-implicit integration scheme with a small time step (see Table \ref{tab:parameters}) to
accurately integrate the stiff lubrication forces \cite{Bian2014, Vazquez-Quesada2016a}.

\subsection{Zero flux boundary condition}
We employ the extra force term on the fluid $\bbf_i^Q$ to impose the zero mass flux along the $z$-direction.
In the Boussinesq approximation the zero mass flux is equivalent to a zero momentum flux, and since the
momentum is trivial to track in the SPH simulations we use $\bbf_i^Q$ to enforce a zero momentum flux along $z$.
In particular, we use a Proportional-Integral (PI) controller
\footnote{That is it, a Proportional-Integral-Derivative (PID) controller without the derivative term.}
\cite{Franklin1998}
\eqn{
  \bbf_i^Q(t^n) = -\corchete{\fr{k_{\text{PID}}}{N_{\text{sph}}\dt} g_z^n + \fr{k_{\text{PID}}}{N_{\text{sph}} \tau_{\text{PID}}} \sum_{m=0}^n g_z^m} \be_z,
}
where $g_z^n$ is the total $z$-momentum at time step $n$ and $k_{\text{PID}}$ and $\tau_{\text{PID}}$ are constants that control the
magnitude of the force, see Table \ref{tab:parameters}.
All the fluid SPH particles $i$ are subject to the same force and
the PI controller adjust the magnitude of the force over time to ensure the average momentum is zero and the deviations small.

\section{Additional results}

\begin{figure}
  \centering
  \includegraphics[width=0.41\columnwidth]{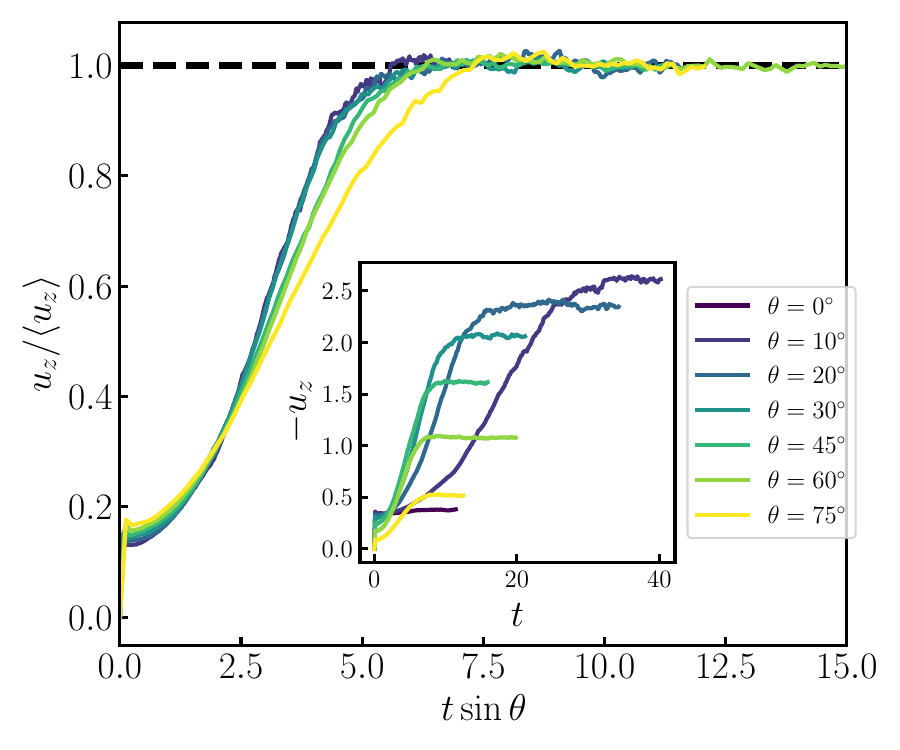}
  \includegraphics[width=0.41\columnwidth]{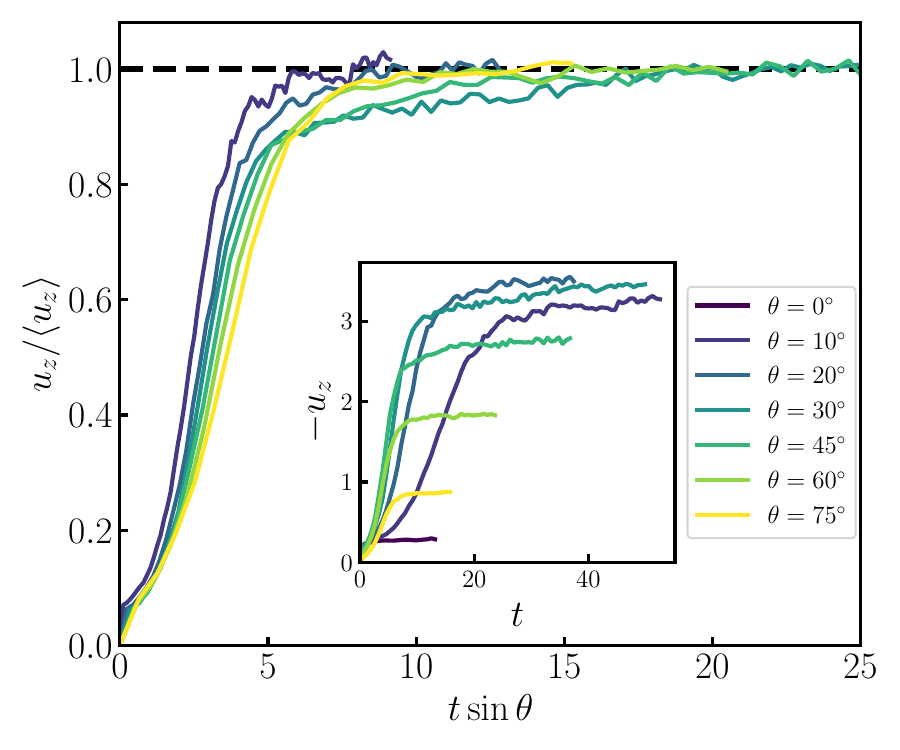}
  \includegraphics[width=0.41\columnwidth]{plot_velocity_vs_time_v4.phi.0.3.longer.py.pdf}  
  \includegraphics[width=0.41\columnwidth]{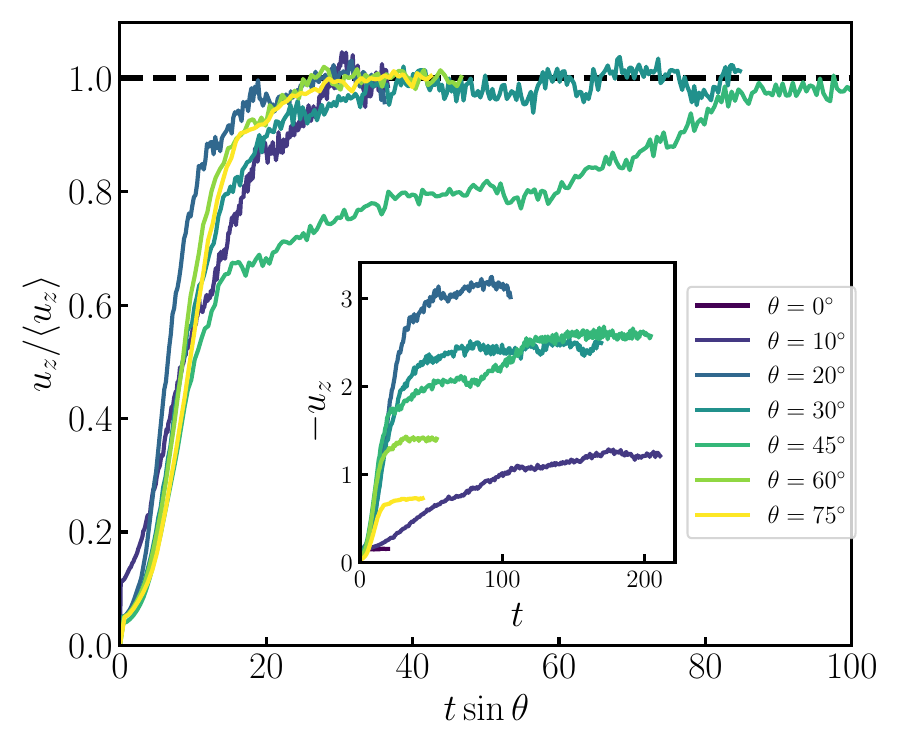}
  \caption{Particle sedimentation velocities versus time for tilts
    $\theta \in [0^{\circ}, 72^{\circ}]$ (curves from dark to light)
    and for volume fractions $\phi=0.11,\, 0.21,\, 0.3 \text{ and } 0.37$ from left to right and top to bottom.
    In the main figures the velocities and the time are scaled with the tilt-dependent steady state velocity $\avg{u_z}$ and with $1/\sin \theta$ respectively.
    The insets show the same data without scaling.
    The curves for $\theta = 0^{\circ}$ are only shown in the insets.
  }
  \label{fig:vel_vs_time}
\end{figure}

The time-dependent sedimentation velocity is shown in Fig.\ \ref{fig:vel_vs_time} for all the volume fractions considered.
For a given channel and volume fraction the curves collapse when the velocities are scaled with the
steady state sedimentation velocity, $\avg{u_z}$, and the times with the inverse of the gravitational force
in the transverse direction $1 / \sin \theta$.

\begin{figure}
  \centering
  \includegraphics[width=0.9\textwidth]{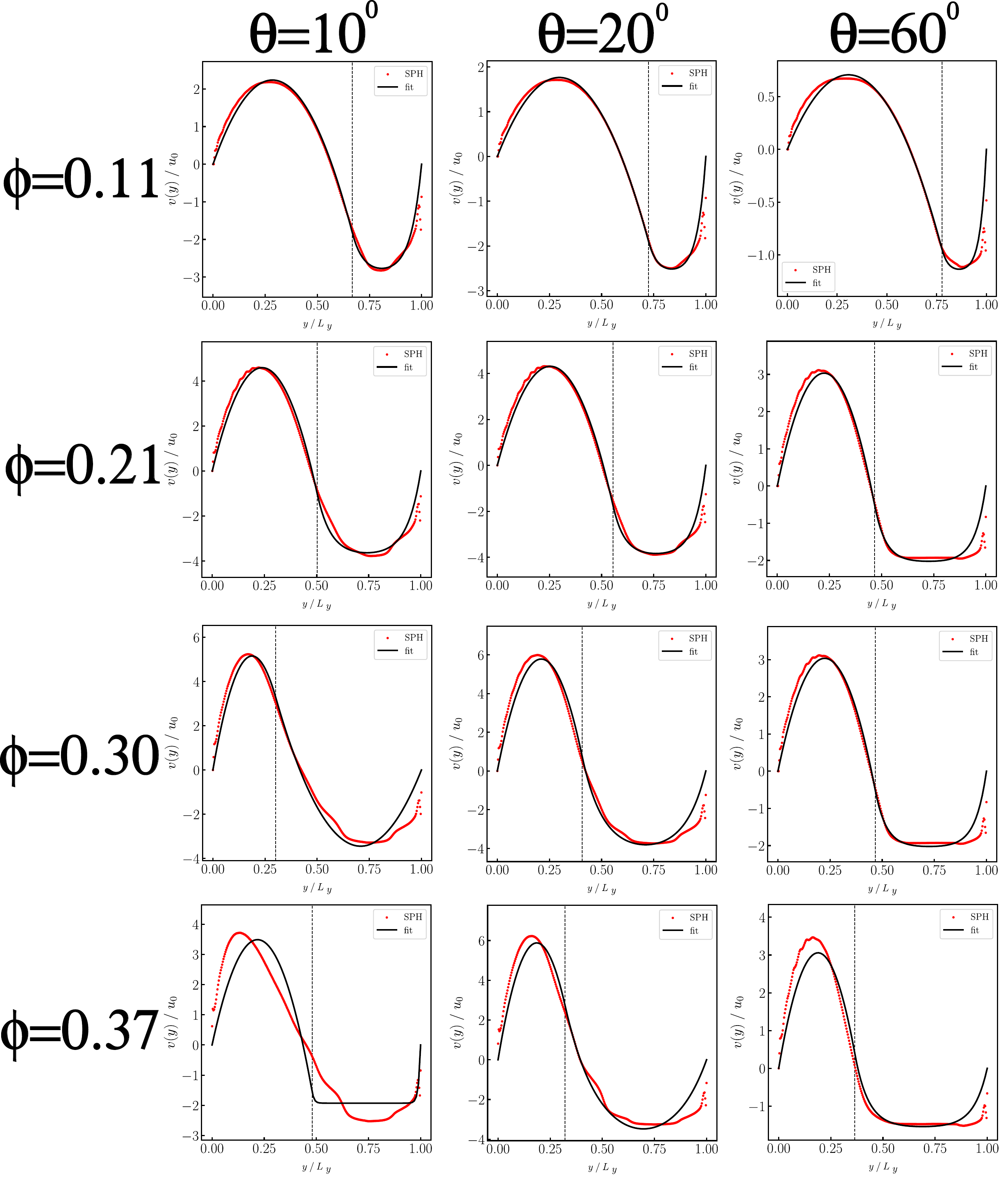}  
  \caption{Velocity profiles normalized with the velocity of an isolated particle, $u_0$, and fits
    to the one-dimensional Stokes Brinkman model (Eqs.\ 6-8 in the main document).
    Columns for for tilts $\theta = 10, 20 \text{ and } 60^{\circ}$
    and rows for volume fractions $\phi=0.11,\,0.21,\, 0.30 \text{ and } 0.37$.
  }
  \label{fig:profile}
\end{figure}

Fig. \ref{fig:profile} shows the velocity profiles across the channel in the steady state and the fits
to the one-dimensional theory discussed in the main document for all volume fractions considered and
several tilt angles.

\begin{figure}
  \centering
  \includegraphics[width=0.32\textwidth]{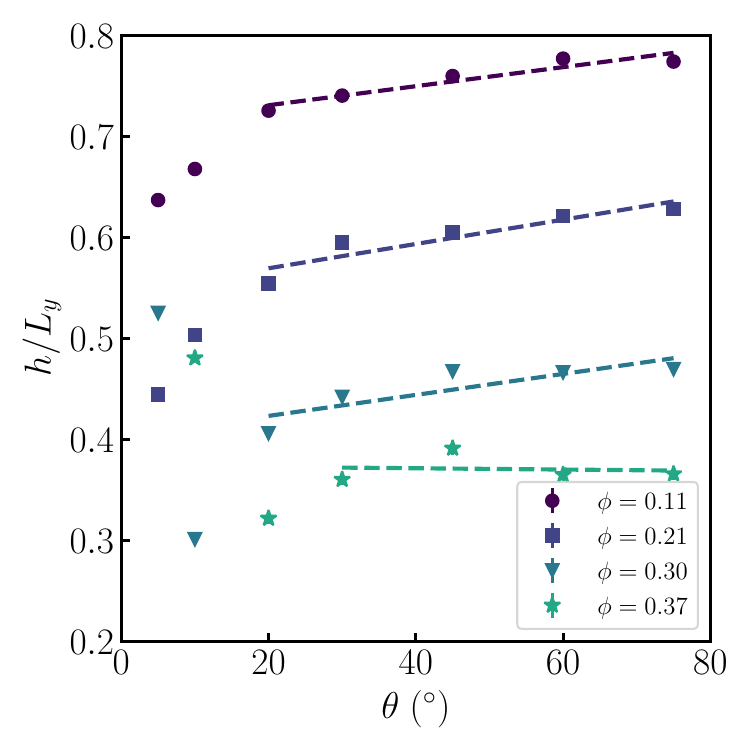}
  \includegraphics[width=0.32\textwidth]{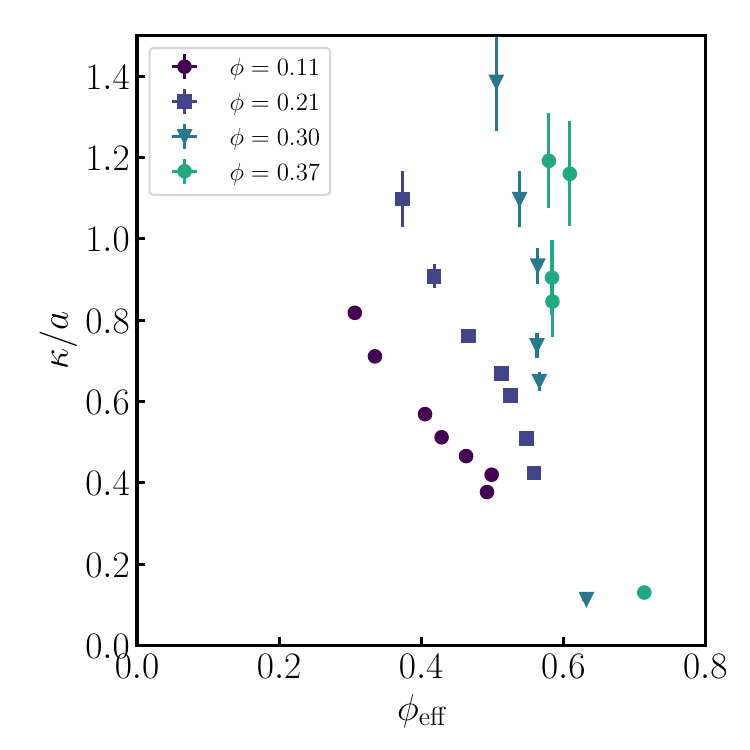}
  \includegraphics[width=0.32\textwidth]{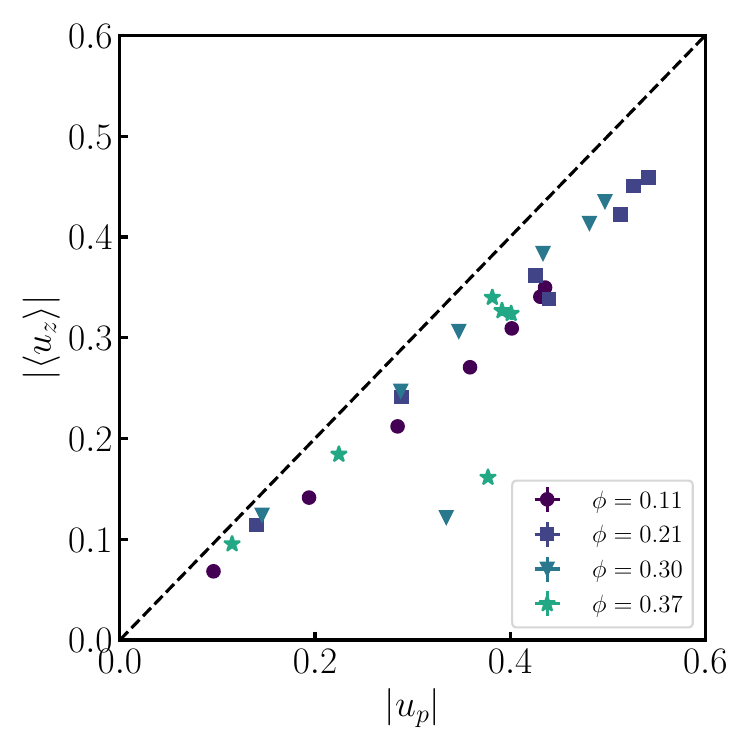}
  \caption{One-dimensional model parameters obtained from the fits of the numerical velocity profiles
    to the one-dimensional Stokes Brinkman model (Eqs.\ 6-8 in the main document).
    {\bf (Left)} Position of the interface $h$ versus tilt angle $\theta$ for several volume fractions.
    {\bf (Middle)} Length scale $\kappa$ (square root of the permeability $\kappa^2$) versus the effective
    volume fraction near the lower wall defined as $\phi_{\text{eff}} = \phi L_y / (L_y - h)$.
    {\bf (Right)} Average sedimentation velocity $|\avg{u_z}|$ versus the velocity of the solid fraction
    $u_p$ of the Brinkman model.
  }
  \label{fig:parameters}
\end{figure}

Fig.\ \ref{fig:parameters} shows the fitting parameters of the one-dimensional model,
$h$, $\kappa$ and $u_p$, obtained from the velocity profile fits.


\begin{figure}
  \centering
  \includegraphics[width=0.4\textwidth]{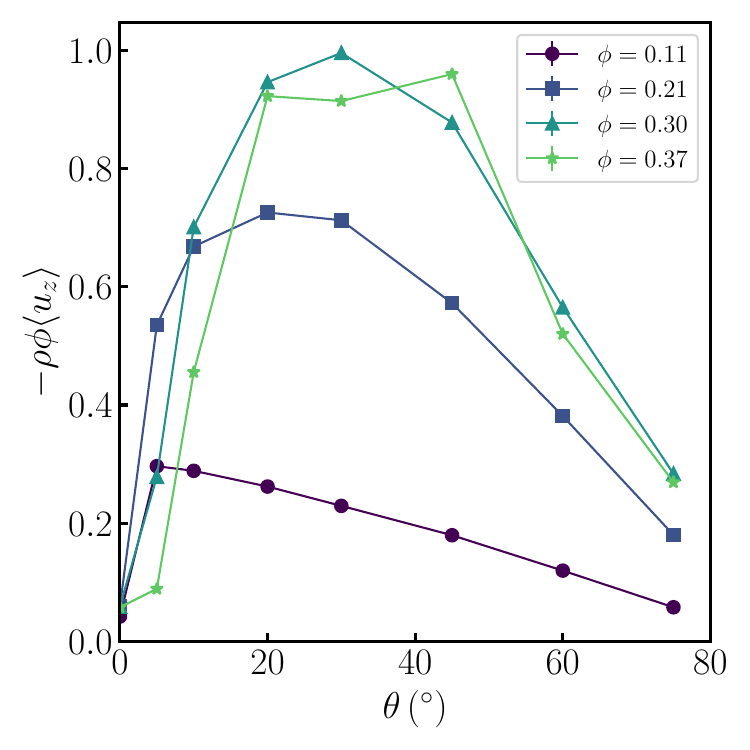}
  \caption{Flux mass, $j_m = \rho \phi \avg{u_z}$, in the steady state versus tilt angle.
    For this system the mass flux is maximize by the tuple ($\phi \approx 0.3$, $\theta \approx 30^{\circ}$).
  }
  \label{fig:mass_flux}
\end{figure}

Fig.\ \ref{fig:mass_flux} shows the mass flux, $j_m = \phi \avg{u_z}$, in the steady state versus tilt angle.

\FloatBarrier
\bibliography{Biblio}